# Online Schema Evolution is (Almost) Free for Snapshot Databases

To appear at VLDB 2023

Tianxun Hu
Simon Fraser University
tha110@sfu.ca

Tianzheng Wang
Simon Fraser University
tzwang@sfu.ca

Qingqing Zhou
Tencent Inc.
hewanzhou@tencent.com

## ABSTRACT

Modern database applications often change their schemas to keep up with the changing requirements. However, support for online and transactional schema evolution remains challenging in existing database systems. Specifically, prior work often takes ad hoc approaches to schema evolution with "patches" applied to existing systems, leading to many corner cases and often incomplete functionality. Applications therefore often have to carefully schedule downtimes for schema changes, sacrificing availability.

This paper presents Tesseract, a new approach to online and transactional schema evolution without the aforementioned drawbacks. We design Tesseract based on a key observation: in widely used multi-versioned database systems, schema evolution can be modeled as data modification operations that change the entire table, i.e., data-definition-as-modification (DDaM). This allows us to support schema almost "for free" by leveraging the concurrency control protocol. By simple tweaks to existing snapshot isolation protocols, on a 40-core server we show that under a variety of workloads, Tesseract is able to provide online, transactional schema evolution without service downtime, and retain high application performance when schema evolution is in progress.

## 1 INTRODUCTION

Multi-versioned concurrency control (MVCC) has been widely adopted by open-source and commercial systems to provide high performance for various database applications. The main benefit is that under MVCC, readers may be allowed to proceed even if there are concurrent and conflicting writers [3, 58]. Almost all the mainstream relational database systems—For example, MySQL [40], PostgreSQL [52], SQL Server [33] and Oracle [41]—implement MVCC to offer snapshot isolation (SI) or repeatable read as the default or recommended isolation level. Many in-memory MVCC protocols have also been proposed to well leverage the abundant parallelism and memory available in modern servers [7, 11, 12, 22, 29, 31, 39, 55].

In addition to high performance for forward processing, modern database applications also require graceful handling of schema evolution to satisfy new requirements, e.g., adding columns, creating tables from existing data and changing constraints. This is typically done with data definition language (DDL) statements such as `CREATE TABLE...AS`, `CREATE INDEX` and `ALTER TABLE`. Internally, the DBMS handles DDL statements by updating database metadata to store the new schema, and examining (e.g., against newly added constraints) and migrating existing table data to conform to the new schema. Data verification and migration during schema evolution can incur massive data movement that may block concurrent data manipulation language (DML) statements. It was common for early systems to necessitate service downtime or maintenance windows for schema evolution [43], often during "quiet hours." With continuous deployment and integration becoming a norm, however, schema evolution can happen quite frequently (e.g., several times a week), requiring reduced or no service downtime and robust error handling with little DBA intervention [10, 42, 43]. This in turn requires DDL statements be executed in a way that is (1) online without blocking concurrent data accesses *and* (2) transactional such that in case a DDL statement fails (e.g., attempting to convert `VARCHAR` that contain "illegal" characters to `INT`) the operation can be safely rolled back to leave the database in a consistent state.

### 1.1 Ad hoc Schema Evolution in MVCC Systems

There have been some attempts to support online and transactional schema evolution in single- and multi-versioned systems [4, 9, 32, 35, 38, 43, 47], but they still fall short in two aspects.

First, existing solutions expose many special, corner cases that need to be carefully handled by OLTP engine and application developers. For example, MySQL does not offer transactional DDL, not all the operations are online/atomic, and concurrent DML is often forbidden [36, 37]. In case a transaction includes any DDL statements, the transaction will be silently committed, posing correctness risks for applications [6]. Some recent work [4] allows DDL operations to be completed lazily without migrating data right away (by doing it in the background), but is limited to compatible schema changes; in case the change introduces incompatibility (e.g., the previous data type conversion case), the user then has to examine the column content in advance to decide whether the DDL statement should be issued as once the metadata (schema) change is done, it is difficult or impossible to be rolled back as a newer version of the application may have already started to use the new schema, leaving certain data being dropped or unavailable. Handling these special cases significantly complicates system design. Moreover, the special cases may change as the DBMS itself is continuously upgraded, further complicating application design.

Second, past solutions often strongly rely on specific DBMS features, some of which are in essence database applications themselves. This may limit the functionality and performance of schema evolution. For example, some systems rely on views [4] and triggers [47] which present performance bottlenecks as they use table-level locking that forbids concurrent updates. If a system does not implement the required features or deviates from the required behaviors, the efficiency of DDL operations can be negatively affected.

Overall, we observe that the key reason for these issues is DDL support is often an after-thought instead of a first-class citizen that is considered at database engine design time. In other words, they are ad hoc solutions with "patches" applied gradually to the DBMS. Also, there is relatively less attention from academia and (in part as a result) real systems often provide inadequate support, making application developers handle DDL operations "in the trenches" (e.g., with external third-party tools [38]) and often describe DDL operations as "dicey" and "dangerous" [50]. Consequently, application developers often try their best to avoid DDL operations, limiting application functionality and end-user experience.

## 1.2 Tesseract: Data-Definition-as-Manipulation

This paper presents Tesseract, a new approach to non-blocking and transactional schema evolution in MVCC systems without the aforementioned limitations.

Tesseract provides bake-in support for online and transactional schema evolution by directly adapting the concurrency control (CC) protocol in MVCC database engines. This is enabled by leveraging two very simple but useful observations: (1) Conceptually, transactional DDL operations can be modeled by "normal" transactions of DML statements that modify the schema and (sometimes optionally) *entire* tables involved. (2) In MVCC systems—thanks to their built-in versioning support—schemas can be stored as versioned data records and be used to participate concurrency control in ways similar to "normal" data accesses for determining table data visibility. When combined, these observations allow us to devise a *data-definition-as-manipulation* (DDaM) approach that supports online and transactional DDL almost "for free" by slightly tweaking the underlying SI protocol *within* the database engine, without having to rely on additional features such as views and triggers.

Like classic catalog management designs [44], Tesseract associates each table (or other resources, such as a database) with a schema record which is stored in a catalog table which in turn has a predefined schema (e.g., table name, constraints, etc.). Using the system's built-in versioned storage support, such schema records are also multi-versioned. This allows us to implement DDL operations as simple as (1) appending a new schema version via standard record update routines and (2) finishing data verification and migration, both in a single transaction that follows the commit and rollback processes of the underlying CC protocol. To access a table, the DML transactions simply follows the standard SI protocol to read the table's schema record version that is visible to the transaction. The schema version record then dictates which data record versions should be accessed by the transaction.

DDaM is straightforward in concept, but realizing it requires careful considerations. Specifically, DDL transactions can be very long by accessing entire tables, and so could block the progress of other transactions, or be aborted due to frequent conflicts with concurrent transactions. It also adds prohibitively high footprint tracking overhead for writes if we were to follow the SI protocol naively. Moreover, concurrent DDL and DML operations must be handled with care to ensure that logically, a DML transaction with work done based on an older schema version never commits after a newer schema has been installed (otherwise subsequent reads would interpret the old content based on a new schema version, leading to potentially wrong results). To this end, in later sections, we propose a relaxed DDaM design that further adapts the SI protocol to allow more concurrency and reduce unnecessary aborts. Relaxed DDaM is the key for Tesseract to achieve online schema evolution without sacrificing much for DML operations.

Although we focus on MVCC in this paper, single-versioned systems could adopt Tesseract if extra (but straightforward) steps are taken to support versioned schema; we discuss possible solutions later. Tesseract can also work with existing lazy migration approaches [4] to support instant deployment of compatible schema changes, which we demonstrate in later sections.

We have implemented and integrated Tesseract with ERMIA [22], a main-memory MVCC database engine, as its schema management and evolution solution. Following past work, we use several representative schema evolution workloads to evaluate Tesseract. On a 40-core server, compared to prior approaches, Tesseract allows the system to evolve database schemas without service downtime or significantly impacting application performance. As we show in detail later, we often observe only up to ~10% of drop for DML operations with DDL operations that involve heavyweight data copying such as adding a column eagerly.

## 1.3 Contributions and Paper Organization

This paper makes four contributions. ❶ We make the key observation that schema evolution can be models as modifying entire tables, to unify DDL and DML handling and propose a simple but useful data-definition-as-manipulation (DDaM) approach for transactional and non-blocking DDL operations. ❷ We show how simple tweaks can be applied to common snapshot isolation protocols to easily and natively support transactional and non-blocking DDL without ad hoc "patches." ❸ Based on DDaM, we build Tesseract to show Tesseract's feasibility and address challenges brought by DDaM. ❹ We compile a comprehensive set of schema evolution benchmarks to evaluate Tesseract and related work. Tesseract is open-source at https://github.com/sfu-dis/tesseract.

Next, we start with the necessary background in Section 2, and give the basic idea and simple tweaks needed for SI to handle DDL natively in Sections 3–4. Section 5 then describes Tesseract in detail and addresses the challenges of DDaM. We cover evaluation in Section 6 and related work in Section 7, before Section 8 concludes.

## 2 MVCC BACKGROUND

In this section, we give the necessary background on MVCC and clarify the assumptions we make throughout the paper.

### 2.1 Database Model

Following past and recent work on memory-optimized MVCC [7, 11, 12, 22, 31, 55], we adopt the MVCC model described by Adya [1] where the database consists of a set of records, each of which is represented by a sequence of totally-ordered versions. An update then appends a new version to the sequence and delete is modeled as a special case of update that appends a tombstone version. Similarly, inserting a record is treated as an update that appends the first valid version for the record. To read a record, the transaction picks a version that is visible to it depending on the isolation level used (described later). To facilitate this, the system maintains a central counter usually implemented as a 64-bit integer [11, 22]. Upon start (or accessing the first record), the transaction reads the global counter to obtain a *begin timestamp*. Upon commit, the transaction atomically increments the global counter (e.g., using the atomic fetch-add or compare-and-swap instruction [20]) to obtain a *commit*

*timestamp* which indicates the transaction's commit order, and is stamped on every version it created. If a version carries a greater commit timestamp, it is created more recently in logical time.

With such a database model, the transaction can pick a visible version based on the set isolation level. Under snapshot isolation (SI) [3], when reading a record, the transaction always "sees" the latest version that was created before its begin timestamp. To update a record, the transaction must be able to see the latest version (i.e., having a begin timestamp that is greater than the latest version's creation timestamp). Under read committed (RC), however, the transaction always reads the latest version, regardless of its begin timestamp. In both cases, transactions follow the first-updater/first-committer-wins protocol [13]: if there is already an uncommitted new version appended to a record by transaction $T$, then subsequent transactions trying to update/delete the same record must abort. That is, only one uncommitted version is allowed per record. In the rest of this paper, we focus on snapshot isolation as it leverages multi-versioning and is widely used and supported in both research prototypes and complete systems [5, 7, 11, 12, 22, 26, 31, 39, 41, 52, 59].

## 2.2 MVCC/SI in Practice and Assumptions

Different implementations exist for the above database model [58]. Without losing generality, we describe the approach taken by ERMIA [22], a memory-optimized database engine; our implementation is based upon it. Each record is uniquely identified by a record ID (RID) that does not change throughout the lifetime of the record, and a collection of records (each of which consists of a sequence of versions) form a table; based on the specific design, record versions could be managed by DBMS-managed heap pages and table spaces, or by a memory allocator in the heap. In either case, a table is represented by an in-memory *indirection array* (also known as the "mapping table" in some systems [11, 28, 30]) that is indexed by RIDs. Figure 1 describes the idea. Each indirection array entry represents a record and contains a pointer to the latest version of the record. Each version also includes a `next` pointer to the next older version of the record, forming a version chain in old-to-new order [58]. Transactions traverse version chains to locate the desirable version to use. To update a record, the transaction generates a new version and atomically inserts it at the head of the version chain, typically by performing a compare-and-swap (CAS) operation on the indirection array entry (with the "new value" being a pointer to the new version, and the "old value" being the entry value it observed upon starting this update). If the CAS succeeded, then the transaction proceeds with its next steps; otherwise, the transaction will abort (and be retried if desired) because a newer version was already installed by a different transaction.

During forward processing, each transaction keeps track of its footprint by maintaining a write set and optionally a read set. Upon commit, the transaction first conducts a pre-commit phase to acquire a timestamp from the central counter and then stamps each new version generated with the commit timestamp, followed by a post-commit phase that persists log records before returning results to the client. The pre-commit phase in essence commits the transaction in-memory. Therefore, for isolation levels higher than RC and SI, additional checks (e.g., phantom protection and serializability) may be performed before post-commit starts. If the transaction does not survive checks before post-commit, it must be aborted.

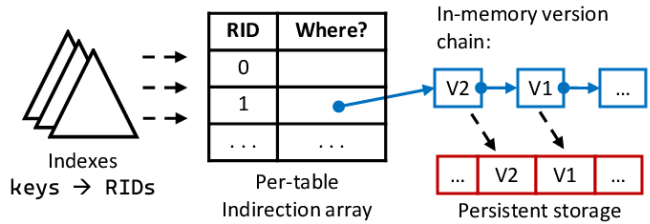

Figure 1: Multi-versioning using indirection. Each table is represented by an indirection array indexed by record IDs (RIDs); each indirection array entry points to a chain of versions of the record. Indexes map keys to RIDs.

Table 1: Examples of DDL operations, categorized by whether they involve copying and/or verifying existing table data. Note that certain operations may appear in more than one category depending on system design (e.g., `ADD COLUMN`).

|           | Copy              | No Copy          |
|-----------|-------------------|------------------|
| Verify    | MODIFY COLUMN     | ADD CONSTRAINT   |
|           | CREATE INDEX      | SET NOT NULL     |
| No Verify | CREATE..AS..SELECT| CREATE/DROP TABLE|
|           | CREATE INDEX      | ADD/DROP COLUMN  |

To avoid log I/O becoming a bottleneck, some systems adopt pipelined commit [21] to decouple log I/O from the commit path, where transactions are queued on a central (or partitioned) commit queue that is monitored by commit/flush thread(s). After pre-commit the worker thread passes the transaction to the commit queue and continues to handle subsequent requests. Once the log records are flushed, the transaction is removed from the commit queue and considered to be fully committed. In memory-optimized systems, thanks to the fact that the entire working set is in-memory, undo logging is typically not required. Therefore, transactions only needs to generate and persist redo logs (i.e., new versions), leading to redo-only logging. However, we note that Tesseract does not depend on whether redo-only or redo-undo logging is used; we only require the availability of replaying log records (described later), which is available in almost all the targeted systems.

## 3 TESSERACT OVERVIEW

The crux of DDaM is to model DDL operations as DML operations that modify entire tables. Certain DDL operations, such as column format conversion from `INT` to `FLOAT`, fit directly with this idea, while others may present different requirements. In this section, we give an overview of Tesseract, by first categorizing the DDL operations to show how different DDL operations can be mapped to DML operations, and then show how the standard SI protocol can be extended with DDaM.

## 3.1 Categorizing DDL Operations

A schema evolution transaction can include a wide range of DDL operations, however, reminiscent of the steal vs. flush decisions in buffer management [44], they largely fall under two dimensions: (1) whether the operation requires actual data copying/modification, and (2) whether the change is limited to verification, i.e., a read-only pass over the table data. We respectively refer to these two dimensions as *copy* and *verify*. A DDL operation then may involve either, both, or none of the two (which indicates the DDL change is limited to the metadata, i.e., "the schema" itself).

Table 1 summarizes several common examples; we omit exhaustive summaries and focus on the common operations to describe the idea and show experiments later. Operations like changing column data type (e.g., from INT to FLOAT using MODIFY COLUMN) involve both "copy" and "verify" as the system needs to scan the table data to ensure the new and old column formats are compatible and then convert it to the new format. In case an incompatible change is detected while the data is being transformed, the DDL operation must be aborted. Certain operations involve only the "verify" dimension. For example, to add a non-NULL constraint to an existing column, one needs to ensure that the entire column does not have NULL values; otherwise the operation is aborted. Similarly, for operations like creating new columns by joining existing tables using CREATE ... AS ... SELECT ... or creating new indexes, the system only needs to generate the new table data or index, without having to verify data types. Finally, it is noticeable that whether a DDL operation involves copy or verify also depends on the underlying system's design and implementation. For example, operations like creating/deleting entire tables could be performed without copy nor verify, by only changing the system catalog (thus listed as no-copy, no-verify in Table 1). Similarly, adding new columns based on a default value could fall under the same category, if the system supports lazy DDL operations: the DDL transaction only needs to modify the table schema to include this new column, along with a function pointer that fills the default value for the field when the data is accesses by subsequent DML transactions.

## 3.2 Schema Versioning

Following classic approaches [44], like "normal" records, a table's schema information is maintained by a "schema record" in a system-wide catalog table which in turn is a normal relational table but with a predefined schema. Under DDaM, this implies that each table is associated with a *schema record* that defines its structure (e.g., data types, columns, and constraints). Figure 2(top) depicts the idea based on the design in Section 2.2. As shown in the figure, the catalog is also represented by an indirection array, and each record in the catalog table is in fact a schema that is also multi-versioned using a version chain. Each schema version also carries a commit timestamp for visibility, along with other information such as the list of columns, constraints and access paths to the data that conforms to it. The schema record's RID also uniquely identifies a data table, i.e., it also serves as the table's unique ID. This way, to access a record, a transaction needs to access a catalog version and the corresponding data version, both subjecting to visibility rules under SI, which we describe next.

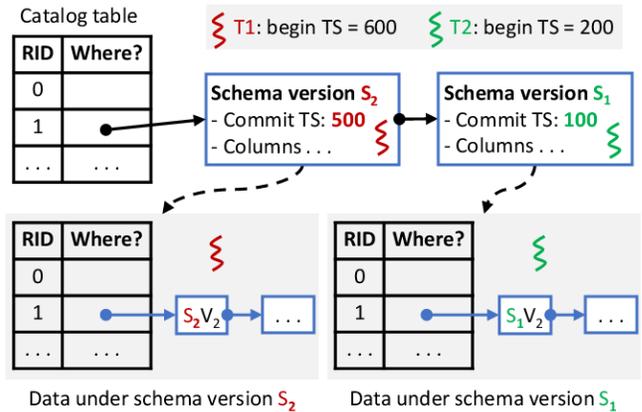

Figure 2: Schema multi-versioning in Tesseract. Schemas are multi-versioned in the same way as table data. Each RID in the catalog uniquely identifies a table. Data under different schema versions is represented by separate indirection arrays; this is not necessary for basic DDaM designs (Section 4) but can extract more concurrency (Section 5). To access a record, the data and schema versions must match: transactions (1) pick the latest visible schema, and (2) read the latest record version that conforms to the schema version.

## 3.3 Transactional DDL and DML Operations

A transaction in Tesseract can include a mix of DML (data record accesses) and DDL (schema evolution) operations. Although the latter in Tesseract are realized using DML operations, for clarity we still refer to them as DDL operations. Next we describe from a high level how Tesseract performs DDL and DML operations.

DML operations still follow the normal SI protocol to pick the suitable version but must ensure the data and schema versions are consistent. As Figure 2 shows, when transaction $T1$ attempts to read a record with RID 1 in Table 1, it visits the catalog table to obtain the schema version, $S_2$, which has a commit timestamp (TS) of 500 that is earlier than $T1$'s begin TS (600). Then, $T1$ uses the schema to access/interpret the latest visible version ($S_2V_2$). However, $T2$ must use $S_1$ and $S_1V_2$ because $S_2$ was created after $T2$ began at TS=200.

Updates are handled similarly, but the protocol should ensure (1) the latest record version *and* (2) the latest schema version are visible to the transaction. The second requirement reflects the key of DDaM: if a concurrent DDL operation has installed a newer schema version but is yet to commit before the DML transaction commits, then the concurrent DDL may be updating the entire table, causing write-write-conflicts and thus aborting either transaction. If the update commits later than the concurrent DDL transaction (but assuming the old schema), the new record version would be stamped a TS greater than the (soon-to-commit) new schema. This will cause later reads to wrongly interpret an older record version using a newer schema, hence must be forbidden.

DDL operations, as we have briefly mentioned, will participate concurrency control just like normal DML transactions, by updating schema records using standard SI protocols, followed by any needed DML operations to transform/verify the table content. The key is that they must carefully handle interactions with DML operations (as described above) and at the same time, leverage the available parallelism in modern multicore CPUs to accelerate the data migration/verification process for it to not become a bottleneck. We provide detailed solutions next.

## 4 DATA-DEFINITION-AS-MODIFICATION

We begin with a basic DDaM design that demonstrates how DDaM can easily enable online and transactional schema evolution; Section 5 then proposes more efficient approaches on top of the basic design. A straightforward approach to DDaM is to model schema accesses as data record accesses by strictly following the generic SI protocols for DML operations. To set the stage, we first review the generic SI protocols, followed by the detailed protocols.

### 4.1 Generic SI Protocols

Based on the high-level ideas in Section 2, Algorithm 1 shows the concrete steps. The vanilla SI protocols are prefixed with generic_ (lines 3–20). In the algorithm (and subsequent ones), we denote tables using indirection arrays indexed by RIDs as described previously. Then, an array entry table[rid] stores a pointer to the latest version on the version chain (following the new-to-old order [58]) for the tuple identified by rid. As shown by lines 3–8 of Algorithm 1, the accessing transaction traverses the version chain to find the latest visible record version along with whether it is the latest. Updating a record as shown by lines 10–20 works by (1) ensuring the transaction can see the latest version (lines 11–13), (2) chaining current head version after the new version (line 16) and (3) issuing a CAS to install a pointer to the new version on the indirection array entry (line 17). If the CAS succeeded, the update is recorded in the write set for commit (described later).

### 4.2 Basic Data-Definition-as-Modification

With the generic SI protocols, reading a schema record is the same as classic approaches that store the catalog as a table, by using generic_read to obtain a visible schema record version (lines 22–23 in Algorithm 1). Here, table IDs are used as "RIDs" in the schema table as we described in Section 3.2.

**Schema Operations.** Under DDaM, the process of evolving a schema requires two steps. First, we update the schema record itself following normal SI protocols: if the latest schema version is not visible to the issuing transaction, it will abort (lines 27–28 of Algorithm 1). After the new schema record version is successfully installed, but *before* the transaction commits, the transaction proceeds to the data migrate phase (if needed) at line 31 which can involve the copy and/or verify actions (Section 3.1) done by generic_read and generic_write. Unlike Figure 2 which in fact shows the "final" Tesseract design (Section 5), the basic DDaM protocol being discussed here does not require a separate indirection array per schema version. Rather, we continue to use a single indirection array per table, which allows data record versions under different schema versions to co-exist in the same version chain; this keeps the design simple but can cause more aborts, which we address in Section 5. After data migration, we proceed to the commit phase (described later). Note that both protocols for reading and updating a schema record are wrapped within a transaction

**Algorithm 1** Vanilla SI protocols for record accesses and schema operations under basic DDaM using the vanilla protocols.

```
1  global: catalog # global schema table

3  def generic_read(t, table, rid):
     # Traverse the data version chain
5    foreach v in table[rid]:
       if v.commit_ts < t.begin_ts:
7        return {<v, is_latest(v)}
     return nil
9
   def generic_write(t, table, rid, new_v):
11   v = table[rid]
     if v.commit_ts > t.begin_ts:
13     return false

15   # Try to install the new version
     new_v.next = v # link new and old versions
17   success = CAS(&table[rid], v, new_v)
     if success == true:
19     t.write_set.add(table.id, rid)
     return success
21
   def ddam_get_schema(t, table):
23   return = generic_read(t, catalog, table.id)

25 def ddam_update_schema(t, table, new_schema):
     # Perform a normal write to install the new schema
27   if !generic_write(t, catalog, table.id, new_schema)
       return false
29
     # migrate data (details in Sec. 4.2-4.3)
31   if !migrate(t, table, new_schema):
       return false
```

context, so they also follow the all-or-nothing atomicity guarantees provided by the engine, enabling transactional schema evolution.

**Data Record Accesses.** Reading or updating a normal data record in a table is straightforward with simple additions of schema accesses before the actual record accesses. Since inserts and deletes are modeled as special cases of updates (Section 2.1), we do not repeat them here. Algorithm 2 shows the full protocol with schema-related steps highlighted. As Section 3.3 describes, to read or write a table record, the only addition to the vanilla SI protocol is to consider the schema version that is visible to the accessing transaction (lines 3–5). Moreover, for write operations, each transaction will also maintain a *schema set* that records all the schema versions used by the transaction (lines 15–16). This is necessary for ensuring commit-time verification, described next.

**Commit Protocol.** After all record accesses are done, the transaction uses ddam_commit shown in Algorithm 2 (lines 19–32) to attempt to commit. Like the vanilla SI protocol, the transaction starts by acquiring a commit timestamp (line 20). It iterates over the schema set to verify that for each updated record, the schema version used by the transaction at the time of the update is still

**Algorithm 2** Tuple read/write and transaction commit protocols under basic DDaM; DDL-related operations are shaded.

```
    def ddam_read(t, table, rid):
2     # Get the schema record version
      {schema, unused} = ddam_get_schema(t, table)
4     if schema == nil:
        return false
6     return generic_read(t, table, rid)

8   def ddam_write(t, table, rid, new_v):
      # Ensure both the latest schema and data are visible
10    {schema, is_latest} = ddam_get_schema(t, table)
      if schema == nil or !is_latest:
12      return false

14    success = generic_write(t, table, rid, new_v)
      if success == true:
16      t.schema_set.add(schema)  # record schema access
      return success
18
    def ddam_commit(t):
20    t.commit_ts = atomic_fetch_add(ts_counter)

22    # Check if the schema matches commit ts
      foreach s in schema_set:
24      latest = schema_table[s.table_id]
        if latest != s:  # a newer schema was added
26        return false

28    # Passed checks, finish the commit
      foreach v in write_set:
30      v.commit_ts = t.commit_ts
        ... same as vanilla SI (persist logs and clean up)...
32    return true
```

the latest at commit time. Note that under SI schema versions for read records are not tracked here. Then, we continue to finalize the new records (including new versions generated as a result of both DDL and DML operations) with the transaction's commit TS at lines 29–31. This is exactly the same as vanilla SI protocols.

### 4.3 Issues with Basic DDaM

By strictly applying the SI protocol, basic DDaM can easily enable online and transactional DDL operations without ad hoc designs. However, strictly following the SI protocol is a double-edged sword that can make DDaM impractical.

We observe several issues with basic DDaM. ❶ DDL transactions are often heavy-weight, long-running transactions as they may incur full-table scans and migration. As Figure 3(a) shows, a long-running DDL transaction may cause other concurrent DML transactions to abort due to write-write conflicts. Note that by definition, transactions that do not conflict with the DDL transaction can proceed in parallel and commit as usual, e.g., the top blue transaction in Figure 3(a). ❷ The DDL transaction itself may get aborted if an earlier concurrent transaction modified a record that happened to be visited later by the DDL transaction. This will waste a lot of useful work done by the DDL and/or DML transactions. Following the first-committer/updater-wins protocol [13] it becomes pure "luck" for schema evolution transactions to successfully commit, as verified by our experiments in Section 6. ❸ Finally, relying on the generic SI protocol means write operations done during data migration will also be tracked. This in turn imposes non-trivial overhead of maintaining a (very large) write set, slowing down commit speed. We address these issues next.

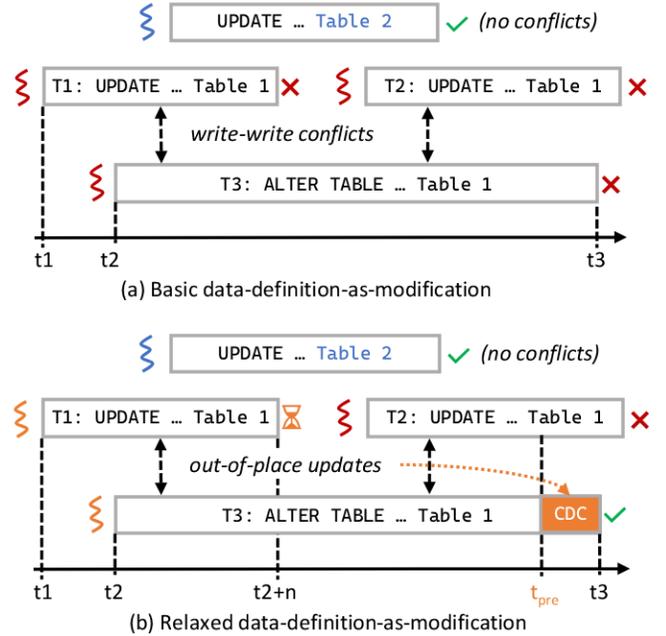

Figure 3: Basic DDaM (a) strictly follows generic SI protocols to easily realize DDL operations, but limits concurrency, causing frequent aborts. Relaxed DDaM (b) allows more concurrency by allowing the new schema to become visible early and pre-committing and verifying concurrent DML transactions while a conflicting DDL transaction is in progress.

## 5 TESSERACT WITH RELAXED DDAM

We observe basic DDaM's issues are largely due to the fact that the generic SI protocol admits very little concurrency among write transactions. First, as mentioned earlier, the DDL transaction can race with a concurrent DML operation to install new versions on the target table using one indirection array, causing aborts upon write-write conflicts. Second, by default a transaction is served by one thread, leading to long conflict windows due to the limited compute capability of a single CPU core. The large write set further exacerbates the situation. We therefore solve the problem by finding ways to (1) allow more concurrent (write) transactions to commit, and (2) increase parallelism while reducing transaction metadata tracking complexity to eventually shorten the conflict window.

Tesseract reaches these goals with relaxed DDaM. Figure 3(b) illustrates the high-level design, which includes (1) an out-of-place migration mechanism, (2) a conflict resolution scheme and (3) relaxed snapshots that still providing the same SI isolation level.

## 5.1 Out-of-Place Migration

Migrating data constitutes the bulk of a DDL transaction if "copy" is needed (e.g., for compatible format conversion). Contrary to basic DDaM in Figure 3(a), relaxed DDaM creates a new indirection array to store new records generated as a result of the DDL operation. This has been shown in Figure 2 where each schema record version also includes a reference to the corresponding indirection array. Until the DDL operation finishes, the new indirection array remains invisible to other transactions. As a result, concurrent DML transactions such as $T1$ and $T2$ in Figure 3(b) continue to use the original indirection array to perform reads and writes, without being aware of concurrent DDL operations. These DML transactions are allowed to proceed and pre-commit internally. For correctness, however, they will not be finalized until the concurrent DDL operation is concluded. We enforce this using pipelined commit [21] mentioned in Section 2.2. In essence, we relax SI's write-write conflict handling protocol to allow more "tentative" writes and defer conflict resolution to a later time via change data capture (Section 5.2). Specifically, in Figure 3(b), $T1$ will be added onto a commit queue (which can be a global figure or partitioned to avoid becoming a bottleneck) and wait. The underlying thread that was handling $T1$ can now switch to handle the next request, without blocking.

Meanwhile, DDL transactions, like $T3$ in Figure 3(b), migrate table data by scanning through the original indirection array using multiple threads. For each record scanned, the DDL transaction (1) transforms the record according to the new schema, and (2) installs the new record version in the new indirection array. While the migration is in-progress, in addition to updates, concurrent DML transactions may also add new records to the original indirection array, potentially increasing the amount of data that has to be scanned. To bound the amount of work done during the scan pass, the DDL transaction captures the size ($S$) of the indirection array upon start, and only scans up to $S$ records in the table. The $[t_2, t_{pre}]$ period in Figure 3(b) indicates this "scan-transform-install" pass.

Since the DDL transaction is the only transaction that can access the new indirection array, the installation step is guaranteed to succeed. However, the transformation step may fail if the intended schema evolution is incompatible with existing data (e.g., to convert a FLOAT column to INT, but the data has decimal points). In this case, the DDL transaction will abort, with all the allocated resources (new record version chains and indirection array, which were never visible to other transactions) reclaimed.[1]

## 5.2 Change Data Capture

After the table is scanned at time $t_{pre}$ in Figure 3(b), the system may have accumulated a series of updates based on the original schema version done by concurrent DML transactions. These updates must be examined and transformed to use the new schema by the DDL transaction. We introduce a change data capture (CDC) phase in Tesseract for this purpose; some systems have been using CDC [25], and Tesseract adapts it for relaxed DDaM. In case of incompatible changes between the update and the new schema, either the DDL or the DML transaction can be aborted, depending on the application's need. If it is desirable to abort the DDL transaction despite wasting much data migration work (e.g., the significant portion of the application code still prefers the old schema), after the "violating" DML transaction commits, subsequent DDL transactions attempting the same change will be aborted as it is incompatible with existing data. In some cases, however, it may be desirable to to abort the violating DML transactions if they are a small stale portion of the application that should conclude soon anyway. Note that aborting the DML transactions is possible because of pipelined commit, but any dependent transactions must also be aborted, causing cascading aborts. Handling cascading aborts may require tracking transaction dependencies, adding extra overhead. Alternatively, one could leverage commit pipelining to avoid tracking dependencies, by simply aborting all the transactions pre-committed after the violating DML transaction. But this may cause innocent transactions to be aborted. Therefore, although our current implementation aborts the DDL transaction, in practice, such tradeoff decisions should be made by considering the application's property.

---
[1]Alternatively, the user may specify resolution approaches in advance, for example, by omitting all the decimal places during the FLOAT to INT conversion. This can avoid the DDL to abort and is orthogonal to our design.

The speed of the CDC phase is critical: if there is too much work, DML transactions intending to use the new schema will wait for a long time, increasing transaction latency. We solve this problem by (1) bounding the amount of CDC work and (2) introducing more concurrency and parallelism for CDC. The CDC phase normally starts after the scan phase, when the DDL transaction has acquired a pre-commit timestamp ($t_{pre}$). The DDL transaction then makes the new schema visible but puts it in a special "pending" state. This way, any new transaction started after $t_{pre}$ should start to use the new schema, thus avoiding new CDC work of data transformation; we discuss the potential of allowing DML transactions under the new schema to use the new indirection array before CDC finishes in Section 5.3. The DDL transaction then continues the CDC phase to scan the log to discover concurrent updates pre-committed up to $t_{pre}$. To facilitate this, the scan phase upon start records the current log sequence number which will be used as the starting point of the CDC phase. The end of the CDC phase, i.e., $t3$ in Figure 3(b), then marks the completion of the schema evolution transaction, unblocking any depending transactions.

With bounded work for CDC, to accelerate the CDC phase itself, we (1) assign multiple threads for CDC, and (2) more importantly, allow CDC to start before the scan phase finishes to shorten the CDC phase. The CDC phase could therefore overlap with the scan phase, but the DDL transaction still pre-commits (i.e., make the new schema visible but pending) only after the scan phase has concluded. Consequently, a CDC thread may conflict with a scan thread of the same DDL transaction. Such conflicts are handled similarly to the SI protocol: both the CDC and scan threads must use a CAS instruction to install new records while observing version ordering such that a version is installed to the version chain only if it is the latest (i.e., the head version is older).

## 5.3 Relaxed Snapshots
Under basic DDaM, the DDL transaction blindly follows the generic SI protocols to participate in concurrency control, migrating the latest visible version based on its own snapshot defined by its begin timestamp. This snapshot can easily become stale as other concurrent transactions proceed with the original indirection array, putting much pressure on CDC later for conflict resolution. Out-of-place migration using a dedicated indirection array further allows us to relax and simplify the SI protocol for both the scan and CDC phases. Instead of strictly following SI protocol by using DDL transaction's begin timestamp to read each record, we allow the DDL transaction to always directly migrate the latest committed version of each record during the scan phase. The newly transformed version on the new indirection array inherits the original record's commit timestamp. This way, the DDL transaction is allowed to migrate versions as fresh as possible, and in fact does not need to maintain a write set for the new versions as the records are marked as committed immediately after migration, greatly reducing metadata tracking overhead of DDL transactions. At a first glance, this design violates the property that a record version always carries a more recent timestamp than its corresponding schema version (Section 4). However, this is correct thanks to the use of a separate indirection array for data migration. First, the newly migrated versions remain invisible until the DDL transaction is fully committed (i.e., after the CDC phase is completed). Second and more importantly, the new indirection array will replace the original one after the schema evolution process finishes. Therefore, any transaction started after the DDL transaction has pre-committed will be using the new indirection array.[2] This in turn means there is only one valid, candidate version that is visible to these transactions. It is then safe to stamp the newly transformed version with the original, smaller timestamps, in exchange for faster DDL commit without having to traverse the write set like the original SI protocol does.

To further reduce blocking caused by CDC, we allow eligible DML transactions to directly proceed without waiting for the DDL transaction to finish. Note if the DDL transaction is related to constraint checking, then all DML transactions should wait the DDL transaction to complete. If the DDL data migration only involves copy operations, DML transactions with blind writes (updates and inserts) on target records can proceed no matter whether those records have been migrated or not. For DML transactions with reads, it is possible to proceed when target records have already been migrated. We identify such records by performing a overlap check: Upon accessing a data record, the transaction simply takes a "sneak peak" of the migrated record on the new indirection array. If such a record does not exist or the record has a commit timestamp that is greater than the one on the original indirection array, then the DML transaction intending to update the record should abort. The reason is that the former case indicates the record is yet to be migrated, while the latter indicates that another DML transaction has already updated the record but again is yet to be migrated. Otherwise, if the new indirection array carries a newer version, the DML transaction is allowed to access the record, but is subject to commit pipelining and abort in case the DDL transaction aborts.

### 5.4 Discussions

Compared to basic DDaM, relaxed DDaM extracts more concurrency by allowing tentative updates from concurrent transactions. These concurrent transactions could cause (heavyweight) DDL operations to abort, however, in practice such cases are very rare. Moreover, as we use more threads to perform DDL operations in parallel to accelerate DDL operations, if the compute resource is on a fixed budget, such cases will be even rarer given most CPU cycles would be put to performing DDL operations, giving higher priority to schema evolution. We thus believe this is a reasonable tradeoff.

Since Tesseract targets snapshot databases, we have focused on SI. DDL transactions in Tesseract participate concurrency control like "normal" DML transactions, so anomalies under SI [3, 14] may also manifest during schema evolution. It is up to the application to carefully analyze the workload and orchestrate concurrent accesses to avoid anomalies, which is the same as what was previously done if the system does not guarantee serializability [13]. Tesseract does not change this behavior, which (as a side benefit) allows application developers to transparently include schema evolution in workload analysis, without taking DDL operations as a special case.

Some systems [4] evolve schemas lazily: once the schema is updated, it becomes visible (committed), while data migration happens in the background or on demand upon record accesses. Tesseract is compatible with and can adopt this approach by further relaxing DDaM to allow early visibility of the new schema, yet without using additional application-level data structures. Nevertheless, with lazy evolution Tesseract will also exhibit the disadvantages seen in other systems. For example, incompatible schema changes could be committed without verification, leaving a diverged database requiring manual investigation by the DBA or application developer. Some DDL operations are eager in essence where lazy evolution would not help (e.g., range index creation mandates the entire DDL operation to finish before any range query can be admitted to avoid missing records). Tesseract directly supports such cases.

Finally, by supporting schema evolution as part of the CC protocol, DDaM and Tesseract make schema evolution much more lightweight. In particular, many existing approaches conduct DDL operations as an "add-ons" using external DBMS features such as triggers [38], i.e., the evolution transaction itself is a user application, leading to high resource consumption. For example, sometimes secondary indexes are still rebuilt, although the indexed column did not change. Tesseract avoids such resource wastes and and certain designs of Tesseract could also be implemented at the application level (e.g., index reusing). Given the wide use of existing solutions, application developers may gradually adopt Tesseract features (e.g., multi-versioned schemas and index reusing, even in the application-level) for a smooth transition.

## 6 EVALUATION

In this section, we empirically evaluate Tesseract using microbenchmarks and variants of standard benchmarks. We compare Tesseract with popular existing approaches and through experiments, we explore the following aspects.

- DDaM and Tesseract support non-blocking transactional schema evolution natively, including a wide range of DDL operations.

---
[2]Pending existing accesses to finish. This can be achieved by an epoch-based memory management scheme or reference count, which many systems already implement.

- Tesseract mitigate the drawbacks of traditional ad hoc approaches to schema evolution.
- Relaxed DDaM can extract much more concurrency among transactions, compared to basic DDaM.

## 6.1 Experimental Setup

We perform experiments on a dual-socket server with two 20-core Intel Xeon Gold 6242R CPUs clocked at 3.10GHz. The server has 40 physical cores (80 hyperthreads) in total and 375GB main memory, and runs Ubuntu 20.04 with Linux kernel 5.8. Worker threads are pinned to physical cores. Unless otherwise specified, we leverage both sockets on the server and observe per-second throughput. All the code is compiled with Clang 10 with all the optimizations.

**Implementation.** We implemented Tesseract in ERMIA [22], an open-source main-memory optimized database engine; our implementation is available at https://github.com/sfu-dis/tesseract. ERMIA already implements snapshot isolation using indirection arrays but did not feature any DDL-related functionality. That is, all the schema information was hard-coded in C++ without a SQL layer. For our evaluation, we implemented a simple catalog manager in ERMIA which is an ordinary table backed by an indirection array like other tables, as described in Section 3.2. On top of these, we then implemented a set of benchmarks (described later) using ERMIA's C++ interface to perform both DDL and DML transactions. As ERMIA is an main-memory engine, all the table data is stored in DRAM. Following previous work [22, 24, 54, 56], we also store log records in DRAM-backed `tmpfs` to stress the system by ruling out the impact of storage I/O.

**Approaches under Comparison.** We run experiments under the following approaches, which are all implemented in ERMIA:

- `NoDDL`: Vanilla ERMIA without any DDL functionality. We use it to show the upper bound.
- `Blocking`: Baseline approach that implements table-level locking for schema consistency. DML transactions which do not intend to evolve schemas acquire the lock in reader mode, while DDL transactions acquire the lock in writer mode. The lock is implemented using `pthread_rwlock_t` [19].
- `Lazy`: A lazy migration modeled after BullFrog [4] that implements DDL functionality at the user level and only updates schema records in DDL transactions, with data migration done in the background or on demand.
- `Tesseract`: Tesseract with relaxed DDaM described in Section 5.
- `Tesseract-Lazy`: Same as `Tesseract` but adapts lazy data migration at the engine level to enable optimizations such as index reuse mentioned in Section 5.4.

We have also tested a variant that uses basic DDaM described in Section 4. However, we observed that almost no DDL transaction can ever commit when data migration is involved. We therefore do not show it in the rest of this section for brevity.

**Methodology and Metrics.** Since schema evolution transactions are typically heavyweight, for each workload, we issue one DDL transaction. We then focus on and report the throughput of DML transactions under various workloads and each of the aforementioned approaches. We focus on the per-second throughput over a period of time, to highlight the impact of concurrent schema evolution on DML transactions.

## 6.2 Benchmarks

We use both microbenchmarks and variants of the TPC-C [53] benchmark in our experiments.

**Microbenchmarks.** We base on the widely used YCSB [8] benchmarks to devise microbenchmarks to stress test the approaches under evaluation. All the transactions are performed on a single table of three 8-byte integer columns that is preloaded with 100 million records. By default, our YCSB implementation includes DML transactions that uniform randomly picks two records to read, and eight records to update. In addition, we introduce two new DDL operations that can be included by a DDL transaction: `AddColumn` and `AddConstraint`. The former adds another 8-byte column to the database table, and the latter adds a constraint to limit the values from the third column to be less a random number. `AddColumn` evaluates the underlying system's behavior when handling schema evolution that involves copying/transforming data, whereas `AddConstraint` stresses "verify" operations as defined by Section 3.1. We then devise three DDL transactions that respectively issue (1) only `AddColumn` (2) `AddConstraint` and (3) both operations.

The benchmark starts with threads keep issuing and finishing DML transactions. Then after a certain period of time, we issue a DDL transaction to perform one of the aforementioned operations. We tuned the number of threads used for DDL operations and unless otherwise specified, we allocate eight threads for the DDL transaction (three for scanning and five for CDC in `Tesseract`) and 30 threads for DML transactions.

**TPC-C with Schema Evolution (TPC-CD).** The TPC-C benchmark [53] models a warehouse wholesale operation and has been the standard benchmark for OLTP. However, all of its five transactions focused on DML operations. To evaluate Tesseract under real OLTP workloads, we therefore follow previous work [4, 34] to extend TPC-C with a set of DDL operations. We refer to the extended TPC-C benchmark as TPC-CD.

Same as in the microbenchmarks, we start the benchmark by running the original TPC-C mix, and after two seconds, we start a DDL transaction to perform one of the following operations:

- `AddColumn`: Add a column `ol_tax` in the `order_line` table with a default value of 0.1.
- `AddConstraint`: For the `order_line` table, add a constraint to require $1 \leq ol\_number \leq o\_ol\_cnt$.
- `AddColumnWithConstraint`: In addition to `AddColumn`, also add a constraint to require $ol\_amount \geq 0$.
- `SplitTable`: Split the `customer` table into two, one containing private customer information such as credit, payment and balance, and the other containing public customer information like state, city, street, etc.
- `Preaggregate`: Sum up values in the `order_line` table where `order_line.ol_w_id = oorder.o_w_id`, `order_line.ol_d_id = oorder.o_d_id` and `order_line.ol_o_id = oorder.o_id`. The results are added as a new column in the `oorder` table.
- `JoinTable`: Join the `stock` and `order_line` tables. This optimizes the `StockLevel` transaction which reads `stock` after scanning the `order_line` table to get out-of-stock items.
- `CreateIndex`: Create a primary index for the `order_line` table.

By default, we use a scale factor of 50 (except for SplitTable which uses 200 warehouses). Like the microbenchmarks, we tuned the number of DDL and DML threads to respectively use 30 and eight threads for DML and DDL transactions (except for `JoinTable` which uses 24 DML threads and 16 DDL threads). Out of the eight DDL threads, three are allocated to the scan phase, and the remaining five are allocated for CDC work.

### 6.3 Performance under Schema Evolution

Our first set of experiments use the microbenchmarks to focus on DDL and DML operations themselves without interference from application logic. As mentioned earlier, we report the per-second throughput for DML transactions while a DDL transaction is running to observe the impact of schema evolution.

**Copy-Only `AddColumn`.** We first examine the results when the concurrent DDL transaction runs `AddColumn` which involves mainly data copying during migration. As Figures 4(top) shows, all the variants start with a small, constant gap with NoDDL until two seconds later, when we start the DDL transaction. This gap shows the amount of overhead of accessing schema information on the normal record read/write paths. After the schema evolution transaction started, Tesseract maintains its performance when schema evolution is in progress, although with a slight drop of up to ~10%. It fully resumes peak performance after the DDL operation is done at around 12-13 seconds. All the other variants exhibit significant drops during schema evolution. Blocking showed practically zero throughput for five seconds because the DDL transaction has exclusive access to the entire table during schema evolution. Its peak performance is also only around 50% of the other approaches due to pessimistic locking overheads. Lazy performs slightly better with a shorter period of significant drop, but continues to incur high overhead for the next 5-6 seconds. The reason is that although it allows the schema update to complete quickly, subsequent DML accesses must then perform data migration if the background threads have not yet migrated the target records. Lazy also builds schema evolution outside the engine and rebuilds new tables and indexes, further reducing performance. This is in contrast to `Tesseract-Lazy` which implements Lazy at the engine level and can reuse indexes.

**Verify-Only `AddConstraint` and Mixed DDL.** Out of the evaluated approaches, only Blocking and Tesseract can support DDL operations that require verification. Lazy and Tesseract-Lazy allow the schema update to be committed immediately but cannot guarantee full correctness without manual intervention in case of incompatible schema changes. We therefore only show results for Blocking and Tesseract in experiments involving verification. The middle and bottom of Figure 4 respectively show the result when the concurrent DDL transaction runs the `AddConstraint` operation and both operations. Similar to the results obtained earlier for `AddColumn`, here we observe also Tesseract is able to maintain high performance while schema evolution is in progress, whereas Blocking inevitably incurs near-zero throughput for the period of schema evolution time since it forbids any DML operations.

**Summary.** The above results show that Tesseract is able to maintain high performance when concurrent DDL operations are in progress, without having to use ad hoc approaches. The results were obtained under a particular table size and thread allocation

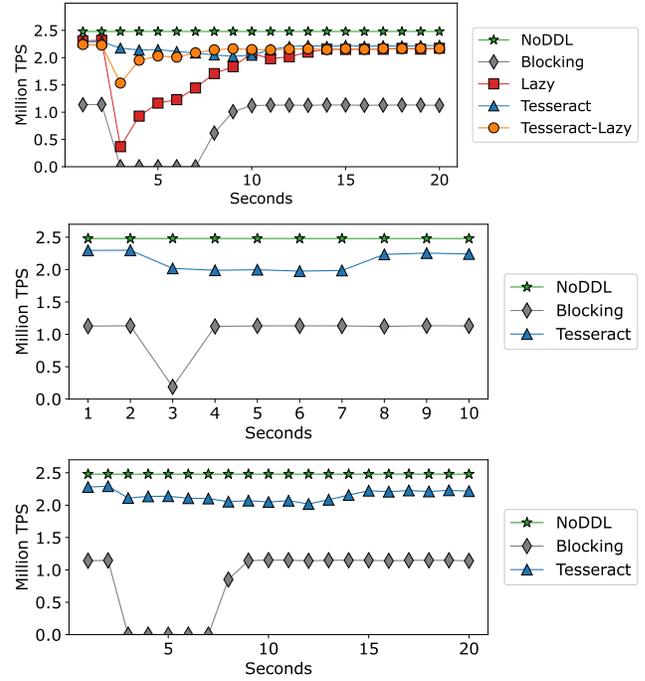

Figure 4: Microbenchmark throughput of DML transactions with a concurrent DDL transaction running `AddColumn` (top), `AddConstraint` (middle) and both (bottom).

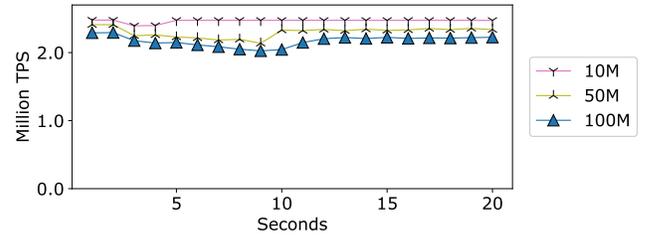

Figure 5: Microbenchmark throughput with a concurrent `AddColumn` DDL transaction under different table sizes.

setup. Next, we explore the impact of these parameters using microbenchmarks.

### 6.4 Impact of Table Size

Now we examine how table sizes affect schema evolution performance. Especially, for copy-dominant DDL operations, a larger table would require more work to be put on migrating records. We repeat the same `AddColumn` experiment done in Section 6.3 but vary table sizes to explore this effect. In Figure 5 we show the throughput under different table sizes from 10 million to 100 million with Tesseract. In general, larger table sizes cause Tesseract to drop slightly more and take longer to finish schema evolution, but without significant impact like other protocols.

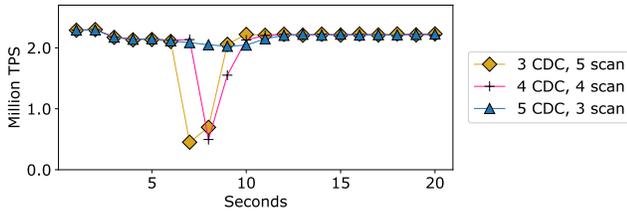

**Figure 6: Microbenchmark throughput with a concurrent `AddColumn` DDL transaction.**

### 6.5 Impact of Worker Thread Allocation

Tesseract employs multiple threads for DDL operations, so it is important to pick the right number of threads for scan and CDC phases for DDL operations to complete quickly. In this experiment, we again run the microbenchmark with the `AddColumn` transaction but vary the number of CDC and scan threads. We keep the total number of DDL threads to eight and vary the CDC/scan threads between 3–5, so that the results are comparable to those reported earlier. As Figure 6 shows, using more CDC threads clearly is the key for Tesseract to maintain its high performance, since when we have 3–4 CDC threads, it becomes hard to keep up with the speed of concurrent DML threads.

Note that this result is obtained under a write-heavy DML workload (80% write), incurring a lot of potential conflicts for CDC to resolve. Therefore, it is beneficial to employ more threads for CDC. However, for read-intensive workloads, fewer CDC threads can be assigned to leave more resources for processing DML transactions.

### 6.6 End-to-End TPC-CD Results

Now we run end-to-end TPC-CD experiments to explore the impact of DDL operations on realistic workloads. We run the standard TPC-C mix and start a DDL transaction of different types after two seconds like in the microbenchmark experiments.

**Copy-Only DDL Operations.** We first explore the impact of DDL operations that require mainly copying during migration. As shown in Figure 7, overall the results showed similar trends to those of the microbenchmarks. As expected, `Blocking` still incurs the highest overhead among all the evaluated approaches. The gaps between `Tesseract` and other non-blocking approaches become smaller as more compute is involved in more realistic workloads.

Under `AddColumn`, Figure 7(a) shows that Tesseract allows smooth migration but is slightly slower than `Lazy` due to the migration process. However, `Tesseract-Lazy` performs better than `Lazy`, showing the potential of Tesseract with lazy migration when the application is sure the intended schema change is compatible with existing data. For the remaining operations (`SplitTable`, `Preaggregate` and `JoinTable`) we observe slightly higher or similar performance for `Tesseract` compared to other state-of-the-art approaches.

**DDL Operations with Verification.** For verification-dominant DDL operations, again we only compare `Blocking` and `Tesseract` for fair comparison. We start with the `CreateIndex` benchmark in Figure 8. In the beginning, the system is barely able to commit any transaction because without the new index, transactions have to issue a full table scan, leading to very low committed transactions per second. After the index creation DDL transaction started the overall TPC-C throughput started to rise quickly under both `Blocking` and `Tesseract`. However, since `Tesseract` allows more concurrency and uses relaxed DDaM to allow early data access concurrent with CDC, it allows more transactions to commit compared to `Blocking`. The `AddConstraint` and `AddColumnWithConstraint` benchmarks in Figures 9–10 show similar trends, with the latter exhibiting a larger gap between `Blocking` and `Tesseract` because its operations are more heavyweight.

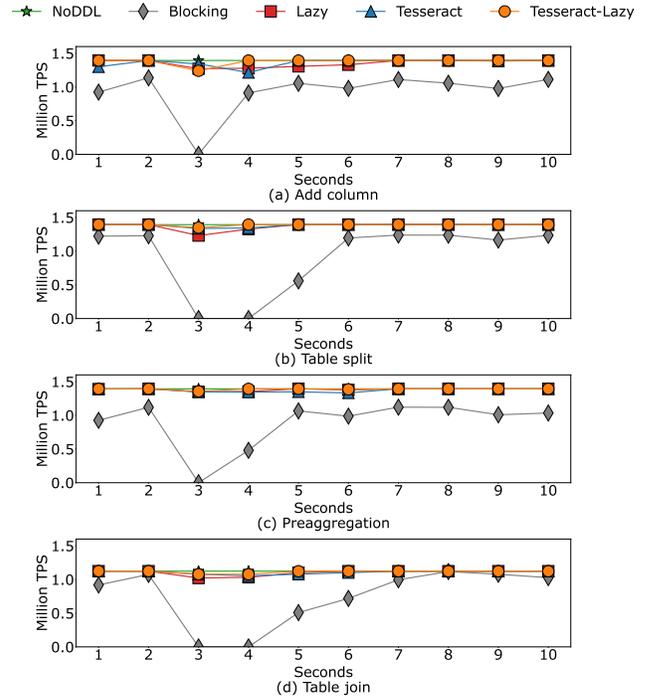

**Figure 7: Throughput of TPC-C DML transactions with a concurrent DDL transaction that mainly migrates data by copying. Most variants except `Blocking` exhibit in general smooth transition. Without ad hoc implementations, `Tesseract` matches the best-performing `Lazy` in `AddColumn` and performs slightly better in other cases.**

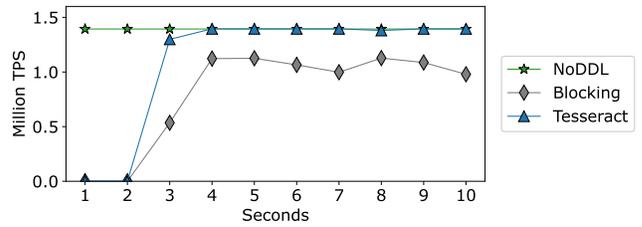

**Figure 8: TPC-C throughput with a concurrent `CreateIndex` DDL transaction.**

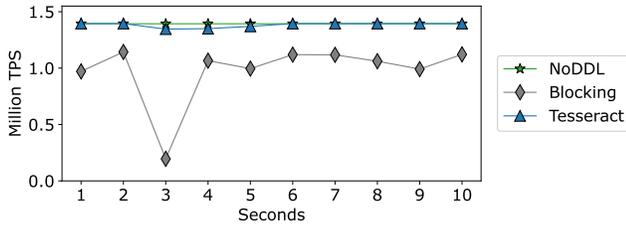

**Figure 9: TPC-C throughput with a concurrent `AddConstraint` DDL transaction.**

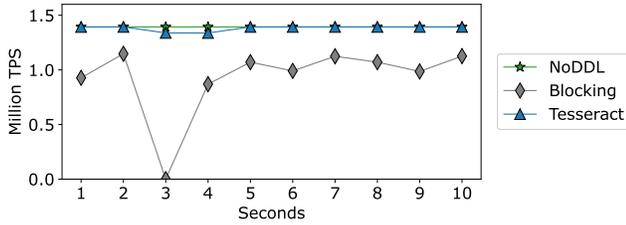

**Figure 10: TPC-C throughput with a concurrent DDL transactions performing `AddColumnWithConstraint`.**

## 7 RELATED WORK

The idea of multi-versioned schema for non-blocking schema evolution is not new and has been discussed in object-oriented [2, 23, 27] and relational [43, 47, 51] database systems. Compared to prior work, Tesseract provides a general, "native" solution that leverages generic SI protocols without relying on ad hoc designs.

**Schema Versioning and Data Migration.** With schema versioning [45], it is important to coordinate data migration such that old and new transactions can co-exist and correspondingly use different schema versions. InVerDa [17, 18] allows co-existing schemas in a system by generating delta code automatically. Changes to data under old schemas are propagated as much as possible to conform to new schemas; incompatible changes can be handled by generating "diffs." Tesseract currently maintains a single lineage of data and the application determines how incompatible changes are handled. Tesseract can be combined with such approaches and can expose interfaces for (read-only) queries of historical data by assigning older read timestamps, because data under different schema versions uses separate indirection arrays. BullFrog [4] migrates data lazily and deploys the latest schema before migrating data on demand or in the background. Compared to Tesseract, BullFrog relies on (thus is limited by) the DBMS's materialized view support, and is mostly applicable to compatible schema changes as incompatible data cannot be migrated at a later time when the new schema is already in use. Similar to soft schema changes [47], Google F1 also allows non-blocking DDL [43] using multi-versioning in distributed systems, but limits the number of schemas to two to simplify maintenance. Sheng [51] proposed a multi-phase lazy migration approach for operations like adding columns, but constraint checking mandates synchronous table scans. KVolve [49] also uses a lazy mechanism for schema evolution, but does not support verification-based DDL operations as constraints are not supported in its targeted systems.

**DDL-Related DBMS Features.** Some DBMS features are necessary or can be helpful for schema evolution. For example, triggers and sagas [15] were used by telecom database applications to perform DDL operations [47]. While sagas can be used to run long migration transactions as a set of short transactions, it is harder to manage. Especially, rolling back involves applying multiple compensating transactions, adding complexity. Similar to our out-of-place migration design, Meta also uses triggers to propagate concurrent writes to a shadow table, which replaces the original table after all data have been migrated [38]. The deferred action framework [60] can support non-blocking schema evolution by registering and deferring them until a later time that is safe to perform DDL operations. Salzberg and Dimock [48] suggests to include all operations in one reorganization transaction for consistency. Wevers et al. [57] used TPC-C to examine the blocking behavior of online schema evolution in MySQL, PostgreSQL and Oracle 11g. Løland and Hvasshovd [32] leverage logging to support schema evolution, but the approach is still based on locking which can limit performance for snapshot databases and is limited in functionality by supporting only full outer join and split transformations.

**Query Rewriting and Languages.** When the schema changes, some systems rewrite user queries to work with a different schema. PRIMA [35] is such a system, but it cannot handle the evolution of integrity constraints. PRISM++ [9] provides better support for handling integrity constraints. Tesseract currently does not use query rewriting, but is orthogonal to and could be combined with such techniques, for example, to allow querying complement provide query answers based on multiple schema versions like PRIMA. Another significant line of work is to provide better query languages for schema evolution [16, 46]. The aforementioned InVerDa [17] system introduces the BiDEL language to generate schema evolution scripts. In comparison, Tesseract works at the OLTP engine level and applications can focus on functionality and business logic.

## 8 CONCLUSION

Transactional and non-blocking schema evolution is an important feature, but is often treated as an afterthought in DBMS design, resulting in ad hoc solutions with extra complexity and sub-optimal performance. In this paper, we advocate for native support for schema evolution within the OLTP engine, by adapting concurrency control protocols. We observe that for snapshot isolation, schema evolution is almost "for free" by modeling DDL operations as DML operations that modify entire tables, leading to our data-definition-as-modification (DDaM) principle. We then proposed Tesseract to adapt snapshot isolation with relaxed DDaM to further allow more concurrency and improve performance. Evaluation results show that Tesseract can maintain high performance while heavyweight DDL transactions are in progress while avoiding ad hoc approaches.

## ACKNOWLEDGMENTS

We thank the anonymous reviewers and associate editor for their constructive feedback. We also thank Qingzhong Meng for discussions and comments. This work is partially supported by an NSERC Discovery Grant, Canada Foundation for Innovation John R. Evans Leaders Fund and the B.C. Knowledge Development Fund.